\documentclass [aps,preprint,showpacs,nofootinbib]{revtex4}
\usepackage{verbatim}
\usepackage{slashed}
\usepackage{epsfig}
\usepackage{graphicx}
\usepackage{amsmath}
\usepackage{mathrsfs}
\usepackage{bm}

\def\prd{Phys.\ Rev.\ {\bf D}}

\def\npb{Nucl.\ Phys.\ {\bf B}}

\begin{document}


\title{QED contribution to the production of $J/\psi+c\bar{c}+X$ at the Tevatron and LHC}


\author{Zhi-Guo He, Rong Li and Jian-Xiong Wang}

\affiliation{Institute of High Energy Physics, Chinese Academy of
Science, P.O. Box 918(4), Beijing, 100049, China, \\Theoretical
Physics Center for Science Facilities, CAS, Beijing, 100049, China.}



\date{\today}

\begin{abstract}
We calculate $\alpha^{2}\alpha_{s}^{2}$ order QED contribution to
$J/\psi$ production in the $pp(\bar{p})\to J/\psi+c\bar{c}$
color-singlet process at the Tevatron and LHC in the framework of
nonrelativistic QCD. The contribution of the interference between
the $\alpha^{2}\alpha_{s}^{2}$ QED and $\alpha_{s}^{4}$ QCD is also
taken into account. The $J/\psi$ production associated with a charm-quark 
pair could be a measurable signal at hadron collider
experiment. Our calculations show that by including the QED contribution,
the $p_{t}$ distribution is enhanced by a factor of
1.5 (1.9) at the Tevatron (LHC) at $p_{t}=50\, (100)$ GeV. In
addition, the polarization of $J/\psi$ turns from unpolarized in all
region to increasingly transverse when $p_{t}$ becomes larger.
\end{abstract}


\maketitle

\section{Introduction}

Heavy-quarkonium production provides important tests on our
understanding of QCD in both perturbative and nonperturbative
aspects. Since the heavy quark mass is larger than $\Lambda_{QCD}$,
heavy quarkonia are commonly thought to be nonrelativistic bound
states. Now the nonrelativistic QCD (NRQCD) effective
theory\cite{Bodwin:1994jh} is widely accepted to study the
heavy-quarkonium phenomenology (for a review
see\cite{Brambilla:2004wf}). In the framework of NRQCD, the
production rates of heavy quarkonium could be calculated with a
rigorous factorization formula based on the double expansions of the
QCD coupling constant $\alpha_{s}$ and heavy quark relative velocity
$v$ in the heavy mesons.

The vital difference between NRQCD and traditional color-singlet
model is that the NRQCD allows the production of heavy quark pair in
color-octet configuration over short distance, and the color-octet state consequently
evolves into the physical heavy quarkonium through the
nonperturbative emission of soft gluons. The introduction of
the color-octet mechanism\cite{Braaten:1994vv} has successfully
reconciled the orders of magnitude discrepancies between the
experimental data measured by the CDF Collaboration\cite{Abe:1997jz}
and the color-singlet model theoretical predictions\cite{j.h.kuhn:79}. However, the  CDF
former\cite{Affolder:2000nn} and latest\cite{Abulencia:2007us}
measurements of $J/\psi$ being almost unpolarized  conflict with the
NRQCD transverse polarization predictions \cite{Beneke:1995yb}. So
the $J/\psi$ production mechanism is still a big
challenge, and various theoretical attempts  can be found in
Refs.\cite{Kramer:2001hh,Lansberg:2008gk,Li:2008ey}.

Recently, substantial progress has been achieved in the
calculation of high order QCD corrections to $J/\psi$
hadroproduction, which are very helpful to clarify the big puzzle.
The next-to-leading-order (NLO) QCD
corrections\cite{Campbell:2007ws} to color-singlet $gg\to J/\psi g$
production enhance the leading-order (LO) cross section by a factor of 2.
Moreover, the NLO result has a large impact on the transverse
momentum ($p_{t}$) distribution of the production rate. It is because
the $p_t$ distribution of the LO result behaves as $1/p_{t}^{8}$,
while the NLO result behaves as $1/p_{t}^{6}$ due to involving the
new topological Feynman diagrams. It is also
found\cite{Artoisenet:2007xi,Hagiwara:2007bq} that the color-singlet
$J/\psi$ production in association with the $c\bar{c}$  pair process
$gg\to J/\psi+c\bar{c}$ also has sizable contribution to the
$\alpha_{s}^{4}$ (NLO) corrections. This process is also
investigated in the $k_T$ factorization formula
\cite{Baranov:2006dh}. Furthermore, it is reported\cite{Gong:2008sn}
that the polarization of $J/\psi$ production via the color-singlet $^3S_1$
state turns from more transverse to more longitudinal after including
the NLO QCD corrections, and most of the $J/\psi$ produced via
color-octet $^{1}S_{0}$ and $^{3}S_{1}$ states at QCD NLO are
still almost in a transverse polarization state\cite{Gong:2008ft} in the large $p_t$ region.
Toward the resolution of the puzzle on $J/\psi$ production mechanism, the study
has been extended to looking for other possible experimental observable
$J/\psi$ production processes. $J/\psi$ production associated with a
photon in the process $pp(\bar{p})\rightarrow J/\psi+\gamma$ is
studied in Ref.~\cite{Drees:1991ig}. Its NLO correction has recently been 
performed in Ref.~\cite{Li:2008ym} and the real
next-to-next-to-leading-order (NNLO) analysis has been 
obtained in Ref.~\cite{Lansberg:2009db}. Double $J/\psi$
hadroproduction has been studied in Ref.~\cite{Barger:1995vx}.

Among these attempts, the study on the $J/\psi$ associated
production, $i.e.$ $J/\psi+c\bar{c}+X$ is very interesting. The
reasons come from two sides. Experimentally, the cross section of
the associated production could be measured by searching for
$J/\psi$ together with $D\bar{D}$ pair, where $D$ denotes all
possible charm hadrons such as $D_{0},D^{+},D^{\ast},\Lambda_{c},$
$etc$. Theoretically, it is noticed\cite{Nayak:2007mb} that at NNLO
the infrared divergence appears due to the soft interaction between
the associated $c$ or $\bar{c}$ quark and the $c\bar{c}$ pair in
charmonium. Hence such processes are nonfactorizable at NNLO. And
the color-transfer enhancement is introduced  in the
double-heavy-quark-pair production process. One persuasive example
is the recent measurements in $e^{+}e^{-}$ annihilation at center-of-mass
energy $\sqrt{s}=10.6$ GeV at $B$ factories for the inclusive and
exclusive charmonium production via double charm-quark pairs. Belle
collaboration first found\cite{Abe:2002rb}
\begin{equation}
\frac{\sigma(e^{+}e^{-}\to J/\psi+c\bar{c})}{\sigma(e^{+}e^{-}\to
J/\psi+X)}=0.59^{-0.13}_{+0.15}\pm0.12,
\end{equation}
which is confirmed by their latest analysis\cite{:2009nj}. The
experimental result is about 6 times larger than the NRQCD
predictions at LO in $\alpha_{s}$ and $v$
\cite{Cho:1996cg}. The problem is resolved by the recent NLO QCD
corrections to inclusive $J/\psi$ production at
$B$ factories\cite{Zhang:2006ay}. Another example is that
the associated production process $J/\psi+c\bar{c}$ is
also found to be important in
$J/\psi$ photoproduction\cite{Qiao:2003ba}.

During the procedure of studying the charmonium production in
$e^{+}e^{-}$ annihilation, it is found that the QED contribution
is sizable since the $J^{PC}=1^{--}$ state $J/\psi$ can be
produced via one photon fragmentation. For example, (a) the cross
section is very large in $J/\psi$ electromagnetic production at
$e^{+} e^{-}$ colliders~\cite{Chang:1997dw}; (b) in $e^{+}e^{-}\to
J/\psi+\eta_{c}$ process, the contribution of the interference
between QCD and QED could enhance that of the QCD result
by $20\%$, although the ratios of the QED contribution to the QCD
contribution is about $1\%$\cite{Braaten:2002fi}; (c) the production
rate of $e^{+}e^{-}\to J/\psi+J/\psi$\cite{Bodwin:2002fk} at LO
is about $3.7$ times larger than that of $e^{+}e^{-}\to
J/\psi+\eta_{c}$\cite{Braaten:2002fi} at $\sqrt{s}=10.6$ GeV,
although the QCD correction to the former process is
negative\cite{Gong:2008ce}; (d) when the center-of-mass energy
$\sqrt{s}>20$ GeV\cite{Liu:2003zr}, the $J/\psi+c\bar{c}$ produced
in $e^{+}e^{-}$ annihilation through two virtual photons will
prevail over that through one virtual photon.

In this situation, it is necessary to examine what will happen in
$J/\psi$ hadroproduction after including the QED contribution. Our
previous work\cite{He:2009cq} has shown that the photon fragmentation process up to QCD NLO is
very important for the $p_{t}$ distribution of the
production and polarization for $J/\psi$ hadroproduction,
though its contribution to the total cross section is negligible
compared to the QCD contribution. As mentioned above, the $J/\psi$
associated production is also a very interesting process. Following
our previous work, we study in detail the QED
effect on $J/\psi+c\bar{c}$ hadroproduction at the Tevatron and LHC in this paper.
The rest of the paper is organized as follows: In Sec. II, we
briefly introduce the basic formulas and input parameters used in
our calculation. In Sec. III, we discuss the QED contribution to the
$p_{t}$ distributions of $J/\psi$ production and polarizations for the associated $J/\psi$
hadroproduction at the Tevatron and LHC. The conclusion is presented
in Sec. IV.

\section{Basic Formulas}

According to the NRQCD factorization approach\cite{Bodwin:1994jh}, the
$J/\psi+c\bar{c}$  production rate in hadron-hadron collisions could
be expressed as
\begin{eqnarray}
\sigma[pp\to J/\psi+c+\bar{c}+X]=\sum_{i,j,n}\int
dx_{1}dx_{2}G_{i/p}G_{j/p} \times
\hat{\sigma}[i+j\to(c\bar{c})_{n}+c+\bar{c}+X]\langle\mathcal{O}^{H}_{n}\rangle,
\end{eqnarray}
where $p$ is either a proton or antiproton. The short distance part
$\hat{\sigma}$ represents the partonic production of $c\bar{c}$ with
quantum number n. $\langle\mathcal{O}^{H}_{n}\rangle$ is the
nonperturbative long-distance matrix element that parametrizes the
transition of the $c\bar{c}$ pair into $J/\psi$.

The QCD contribution to $pp(\bar{p})\to J/\psi+c\bar{c}+X$ starts at
$\alpha_{s}^{4}$ order with two independent partonic processes:
$gg\to c\bar{c}[^{2S+1}L_{J},\underline{1(8)}]+c+\bar{c})$ and
$q\bar{q}\to c\bar{c}[^{2S+1}L_{J},\underline{1(8)}]+c+\bar{c}$,
where q represents all possible light quarks u,d,s.
$c\bar{c}[^{2S+1}L_{J},\underline{1(8)}]$ denotes the specific state
of $c\bar{c}$  with total spin $S$, orbital angular momentum $L$,
total angular momentum $J$ and 1 (color-singlet) or 8 (color-octet). The
contributions of $J/\psi$ production via the color-singlet
$c\bar{c}[^3S_1,\underline{1}]$, color-octet
$c\bar{c}[^3S_1,\underline{8}]$, and $c\bar{c}[^1S_0,\underline{8}]$
states in the $gg$ fusion processes have been considered in
Refs.\cite{Artoisenet:2007xi,Hagiwara:2007bq,Artoisenet:2008tc}. Since
only the $c\bar{c}[^3S_1,\underline{1}]$ state can directly couple
with one photon, which may bring kinematic enhancement, in this
work, we consider the QED contributions from two processes:
\begin{eqnarray}
g(k_1)g(k_2)\to c\bar{c}[^3S_1,\underline{1}](p_1)+c(p_2)+\bar{c}(p_3),\nonumber\\
q(k_1)\bar{q}(k_2) \to c\bar{c}[^3S_1,\underline{1}](p_1)+c(p_2)+\bar{c}(p_3).
\end{eqnarray}

The amplitude at parton level is expressed as
\begin{eqnarray}
\mathcal{M}\left[i(k_1)j(k_2)\to
c\bar{c}[^3S_1,\underline{1}](p_1)+c(p_2)+\bar{c}(p_3)\right]=\sum_{s_1,s_2}\sum_{3k,\bar{3}l}
\langle
s_1;s_2|1S_z\rangle\nonumber\\
\langle3k;\bar{3}l|1\rangle\times\mathcal{M}\left[i(k_1)j(k_2)\to
c_{k}(\frac{p_1}{2},s_1)+\bar{c}_{l}(\frac{p_1}{2},s_2)+c(p_2)+\bar{c}(p_3)\right],
\end{eqnarray}
where $i,j$ are gluon or quarks,
$\langle3k;\bar{3}l|1\rangle=\delta_{kl}/\sqrt{N_c}$ and $\langle
s_1;s_2|1S_z\rangle$ are $SU(3)$-color and $SU(2)$-spin
Clebsch-Gordon coefficients for $c\bar{c}$ pair projecting onto the 
color-singlet $^3S_1$ state. At LO in $v$, the projection
operator of Dirac spinor is\cite{Kuhn:1979bb}:
\begin{equation}
P_{1,S_z}(p_1)\equiv\sum_{s_1,s_2}\langle s_1;s_2|1S_z\rangle
v(\frac{p_1}{2},s_1)\bar{u}(\frac{p_1}{2},s_2)=\frac{1}{2\sqrt{2}}
\slashed{\epsilon}(S_z)(\slashed{p}_1+2m_c).
\end{equation}

To obtain the polarization distribution, the polarization vectors
$\epsilon^{\mu}(\lambda)$ are kept explicitly during our
calculation. The partonic differential cross section for polarized
$J/\psi$ can be written as
\begin{equation}
\frac{\mathrm{d}\hat{\sigma}}{\mathrm{d}p_t}=A\epsilon(\lambda)\cdot\epsilon^{\ast}(\lambda)+
\sum_{i,j=1,2}A_{ij}p_{i}\cdot\epsilon(\lambda)p_{j}\cdot\epsilon^{\ast}(\lambda),
\end{equation}
where $\lambda=T_{1},T_{2},L$. $\epsilon(T_1)$, $\epsilon(T_2)$ and
$\epsilon(L)$ are the two transverse polarization vectors and the
longitudinal one for $J/\psi$, respectively. $A$ and $A_{i,j}$ are the
coefficients. The polarization parameter $\alpha$ is defined as
\begin{equation}
\alpha(p_t)=\frac{\mathrm{d}\sigma_{T}/\mathrm{d}p_t-2\mathrm{d}\sigma_{L}/\mathrm{d}p_t}
{\mathrm{d}\sigma_{T}/\mathrm{d}p_t+2\mathrm{d}\sigma_{L}/\mathrm{d}p_t}.
\end{equation}
We chose physical gauge for gluons to avoid computing the ghost
diagrams. The gauge invariance is checked to ensure the validity of
our results by replacing the gluon polarization vector with its
momentum in numerical computation. Moreover, the long-distance
matrix element $\langle\mathcal{O}_{1}^{J/\psi}\rangle$ is estimated
from the leptonic decay of $J/\psi$ using the relation
\begin{equation}
\langle\mathcal{O}_{1}^{J/\psi}\rangle=\frac{2N_c(2J+1)|R_{J/\psi}(0)|
^{2}}{4\pi}.
\end{equation}
At NLO in $\alpha_{s}$ and LO in $v$, we have $\Gamma(J/\psi\to
e^{+}e^{-})=\frac{4\alpha^{2}}{9m_{c}^{2}}(1-16\alpha_{s}/3\pi)|R_{J/\psi}(0)|
^{2}$. We employ the Feynman Diagram Calculation package\cite{Wang:2004du} to
generate the Feynman diagrams and calculate the invariant amplitudes
numerically with the basic formulas mentioned above.

The $pp(\bar{p})\to J/\psi+c\bar{c}$ process is part of the QCD real
corrections to the $\alpha_{s}\alpha^{2}$ order $J/\psi$ inclusive
production. So we chose the same set of numerical inputs as our last
work with
\begin{itemize}
\item [$\bullet$]QCD coupling constant: $\alpha_{s}(M_z)=0.118$;
\item [$\bullet$] renormalization scale $\mu_r$ and factorization scale $\mu_f$: $\mu_r=\mu_f=\mu_{0}=\sqrt{(2m_c)^2+p_{t}^2}$;
\item [$\bullet$]QED coupling constant: $\alpha=1/128$;
\item [$\bullet$]charm-quark mass:       $m_c=1.5$ GeV;
\item [$\bullet$] $J/\psi$ leptonic decay width: $\Gamma(J/\psi\to
e^{+}e^{-})=5.55\mathrm {keV}$\cite{Amsler:2008zz};
\item [$\bullet$] long-distance matrix element
$\langle\mathcal{O}_{1}^{J/\psi}\rangle=1.35 \mathrm{GeV}^3$ from the leptonic decay width;
\item [$\bullet$] PDF set:  CTEQ6M\cite{Pumplin:2002vw};
\item [$\bullet$] kinematic cut: $p_t>3GeV$; and $|y_{J/\psi}|<0.6$
(Tevatron),
$|y_{J/\psi}|<3.0$ (LHC).
\end{itemize}

\section{QED Contribution to $pp(\bar{p})\to J/\psi+c\bar{c}$}
\begin{figure}[!htbp]
 \begin{center}
  \includegraphics[width=0.90\textwidth]{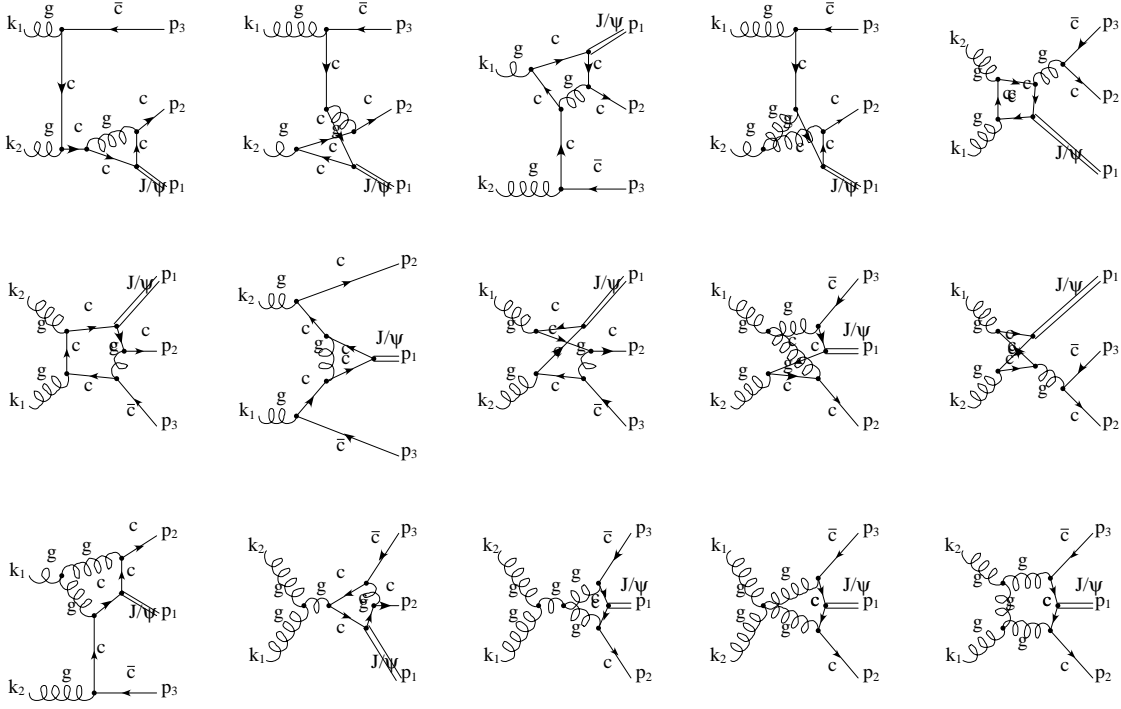}
    \caption{The typical Feynman diagrams for the $\alpha_{s}^{4}$ QCD part of $gg\to
c\bar{c}[^3S_1,\underline{1}]+c+\bar{c}$.}
   \label{fig:QCDdiag}
 \end{center}
\end{figure}

\begin{figure}[!htbp]
 \begin{center}
  \includegraphics[width=0.90\textwidth]{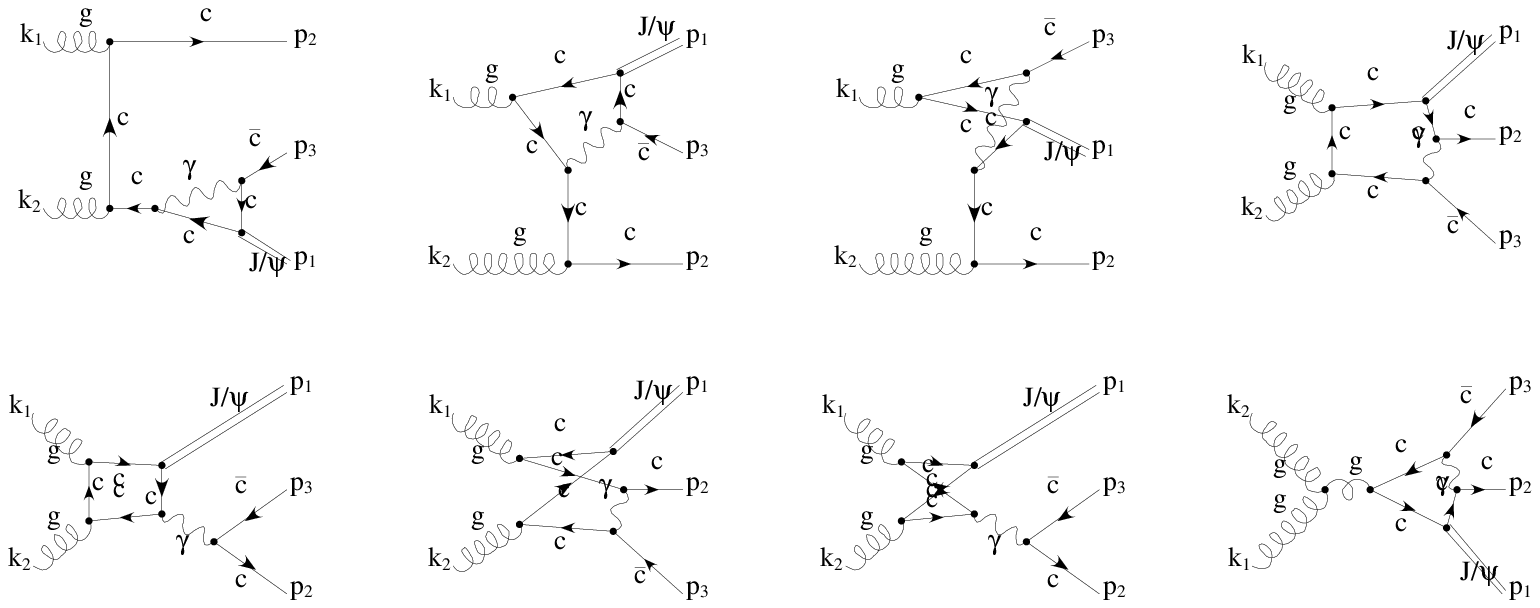}
    \caption{The typical Type I Feynman diagrams in the 
     LO ($\alpha^{2}\alpha_{s}^2$) QED part of $gg\to c\bar{c}[^3S_1,
     \underline{1}]+c+\bar{c}$.}
   \label{fig:QEDdiagI}
 \end{center}
\end{figure}

\begin{figure}[!htbp]
 \begin{center}
  \includegraphics[width=0.90\textwidth]{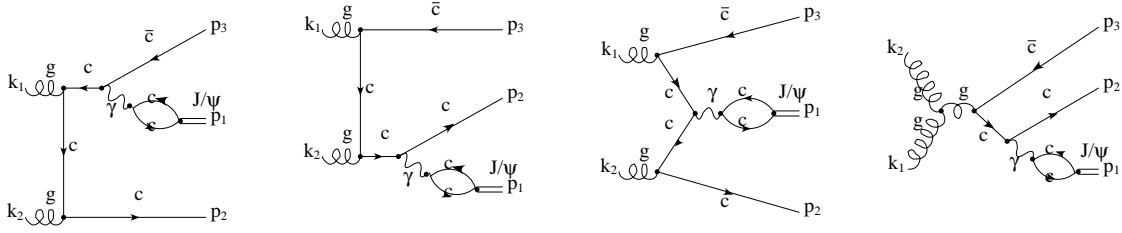}
    \caption{The typical Type II Feynman diagrams of photon fragmentation in the 
     LO ($\alpha^{2}\alpha_{s}^2$) QED part of $gg\to c\bar{c}[^3S_1,
     \underline{1}]+c+\bar{c}$.}
   \label{fig:QEDdiagII}
 \end{center}
\end{figure}

\begin{figure}[!htbp]
 \begin{center}
  \includegraphics[width=0.90\textwidth]{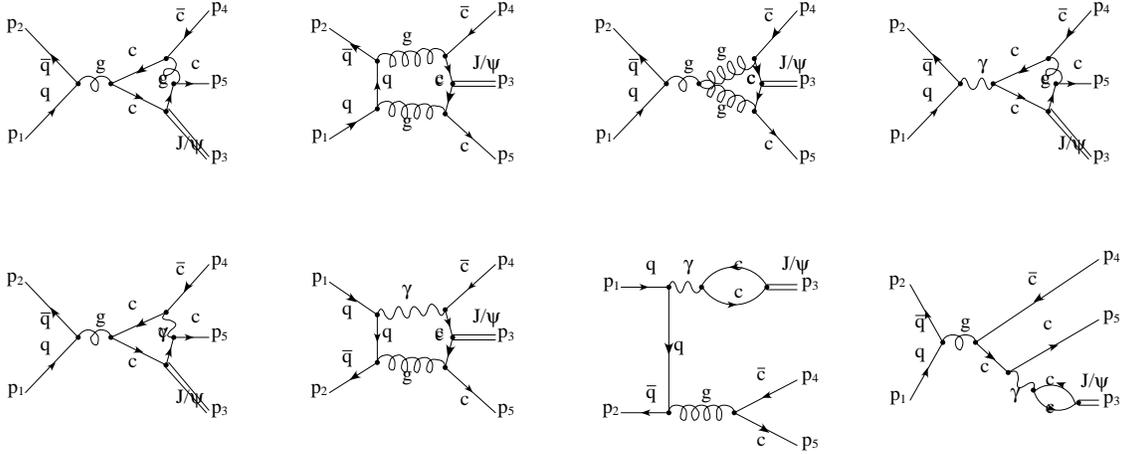}
    \caption{The typical Feynman diagrams of the 
     LO process of $q\bar{q}\to c\bar{c}[^3S_1,
     \underline{1}]+c+\bar{c}$ for QCD and QED.}
   \label{fig:qqbdiag}
 \end{center}
\end{figure}

To investigate the QED effect in $pp(\bar{p})\to J/\psi +c\bar{c}$,
we present the calculation in three parts : QCD, QED and their
interference. The $\alpha_{s}^{4}$ QCD part of $gg\to
c\bar{c}[^3S_1,\underline{1}]+c+\bar{c}$ includes 42 Feynman
diagrams shown in Fig.~\ref{fig:QCDdiag}. While there are only 38 Feynman diagrams for
the $\alpha^{2}\alpha_{s}^2$ QED part, which are divided into two
types and each type forms a gauge invariant subset. The Type I
diagrams shown in Fig.~\ref{fig:QEDdiagI} can be obtained by replacing the
gluon lines between quarks in the QCD diagrams by photon lines.
The Type II diagrams shown in Fig.~\ref{fig:QEDdiagII} are similar with
some of the gluon fragmentation diagrams for color-octet
$c\bar{c}[^3S_1,\underline{8}]+c+\bar{c}$ production when $\gamma^{\ast}\to
c\bar{c}[^3S_1,\underline{1}]$ is changed into $g^{\ast}\to c\bar{c}[^3S_1,\underline{8}]$.
So the partonic differential cross sections of the two types QED
diagrams have the same $1/p_{t}^{4}$ scale behaviors as the QCD
ones\cite{Artoisenet:2008tc}. Then the contribution of Type I
diagrams will be suppressed by $(\alpha/\alpha_{s})^{2}$ compared to
the color-singlet QCD result in both the small and large $p_{t}$
region. However, there is a kinematic enhancement in the Type II
diagrams. The reason lies simplify in the fact that the virtuality of
the photon in Type II diagrams are fixed to $4m_{c}^{2}$, whereas it
varies from $4m_c^2$ to $p_t^{2}$ order in Type I diagrams. When
$p_{t}$ is large enough, the kinematic enhancement may compensate
the suppression factor $(\alpha/\alpha_{s})^{2}$. Furthermore,
the value of $\alpha_{s}(\mu_{0})$ will become smaller as $p_{t}$
increasing. We verified that the contribution of
Type I diagrams is dominant in small $p_{t}$ region and drops fast when $p_{t}$
increases. For the $q\bar{q}$ process, there are 7 QCD and 16
QED Feynman diagrams shown in Fig.~\ref{fig:qqbdiag}, and the QED Feynman
diagrams also can be divided into two types.

\begin{figure}
\begin{tabular}{cc}
\epsfig{file=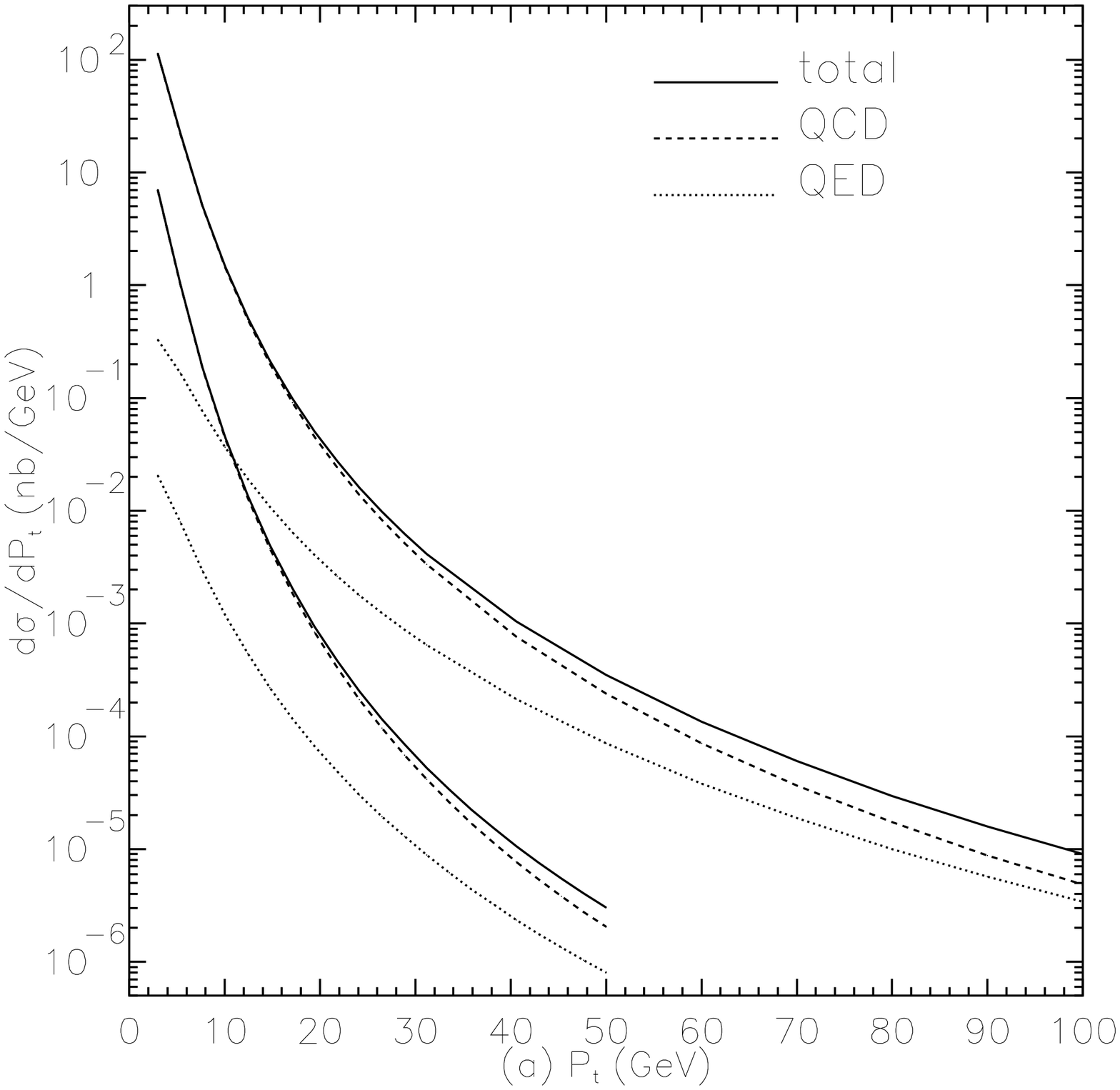,width=7cm}&
\epsfig{file=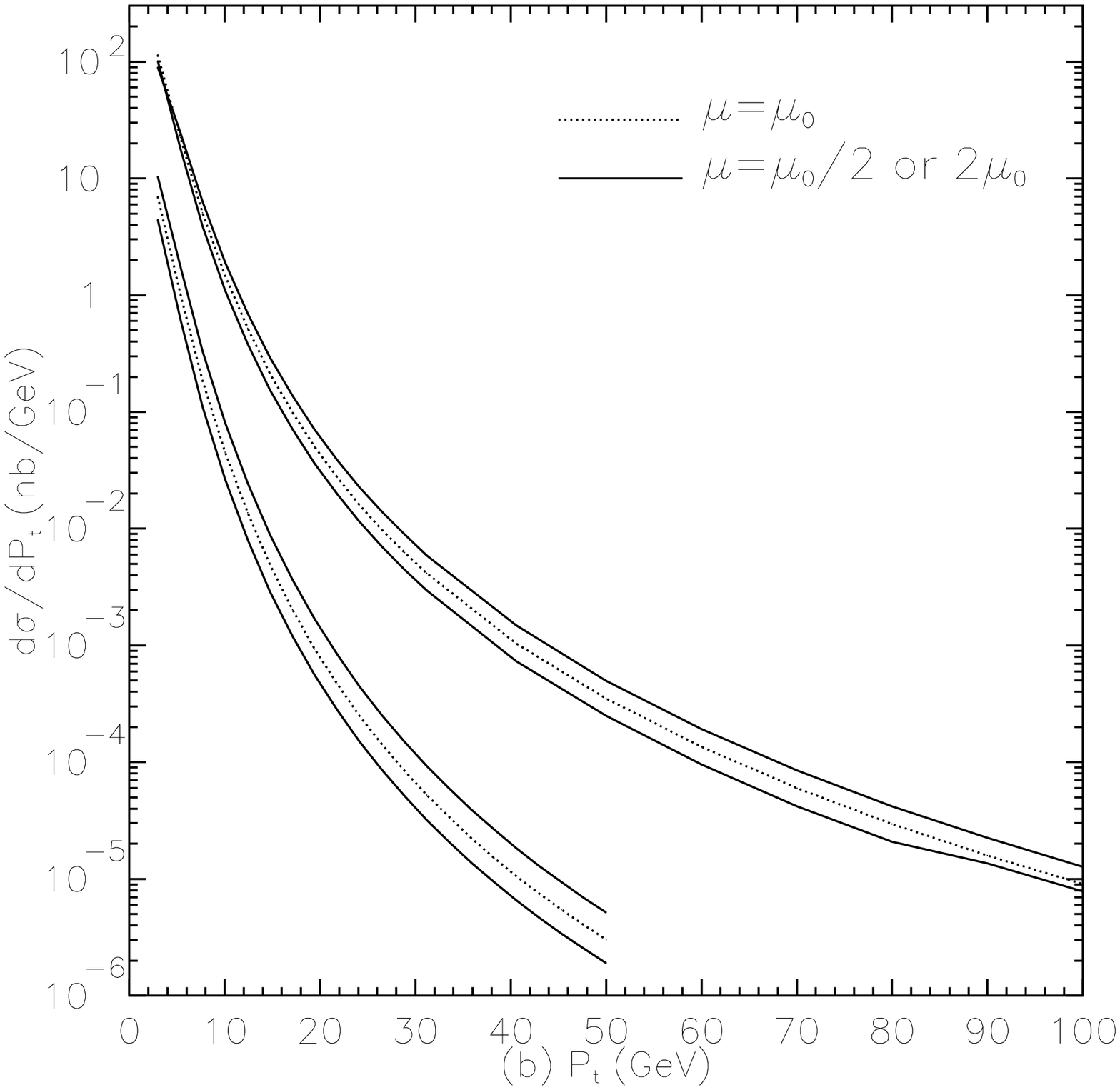,width=7cm}\\
\end{tabular}
\caption{(a): The $p_t$ distributions of the production rate for
$J/\psi + c + \bar{c}$ hadroproduction at the Tevatron ($\sqrt{s}=$1.96TeV, $|y|<$0.6,
3GeV$<$pt$<$50GeV) and LHC ($\sqrt{s}=$14TeV, $|y|<$3, 3GeV$<$pt$<$100GeV).
(b): the $p_t$ distributions of the ``total" with $\mu_r=\mu_f=\mu$ dependence.
Where the total refers to the sum of the QCD, QED and their interference results.
}
\label{fig:ptdis}
\end{figure}

\begin{figure}
\begin{tabular}{cc}
\epsfig{file=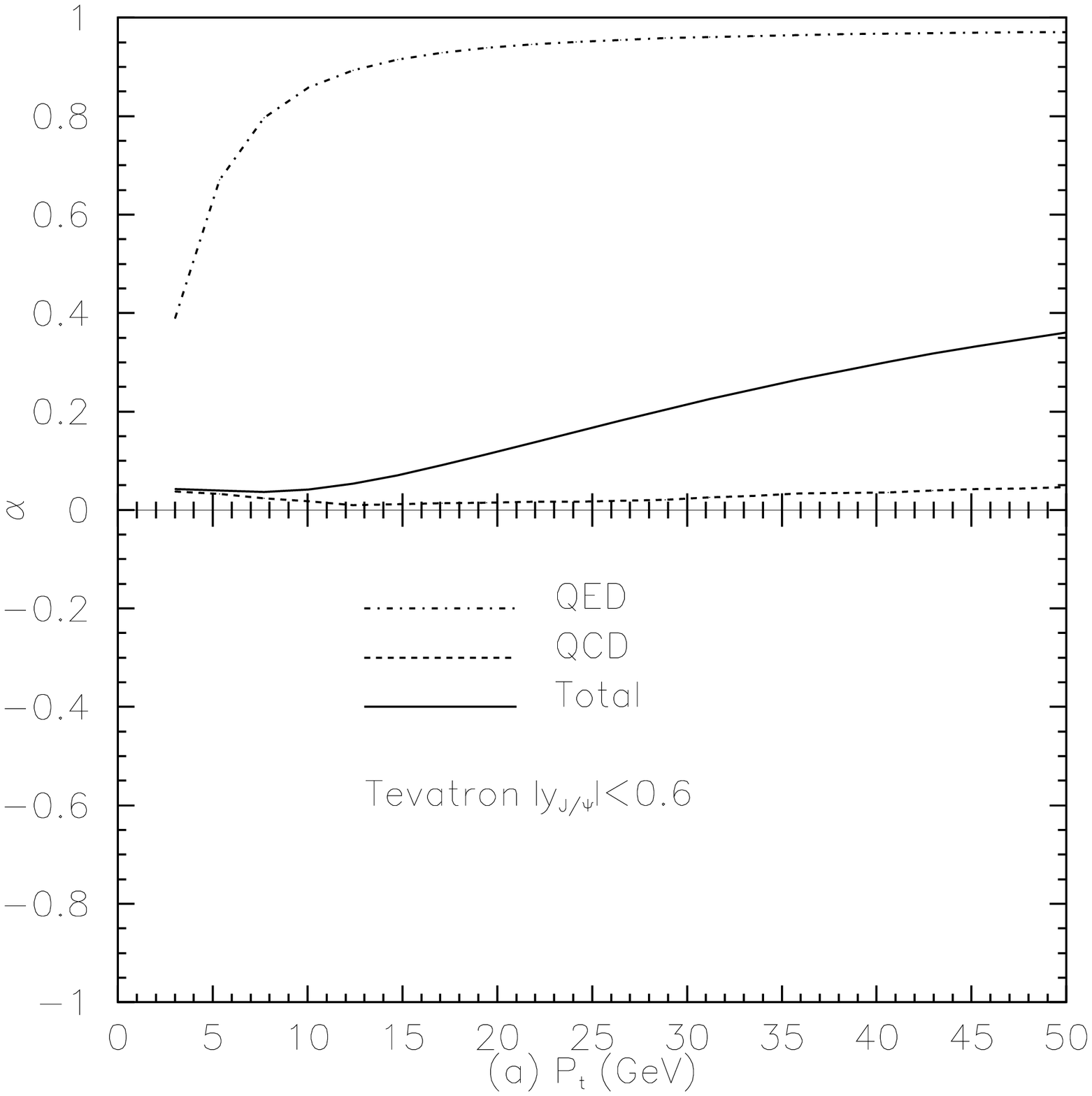,width=7cm}&
\epsfig{file=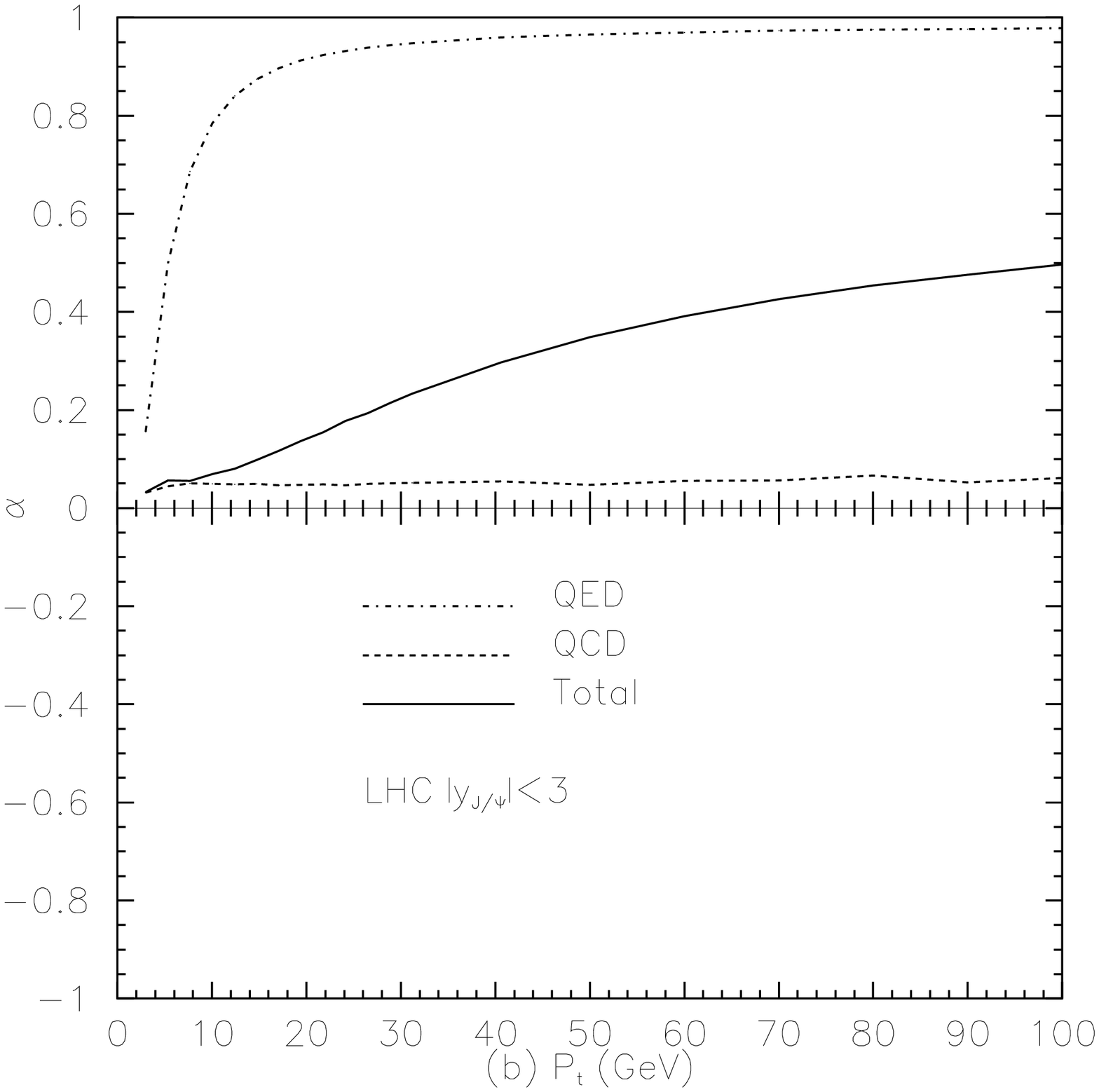,width=7cm}\\
\end{tabular}
\caption{The $p_t$ distribution of polarization parameter $\alpha$ for the 
$J/\psi + c + \bar{c}$ hadroproduction at the Tevatron (a) and LHC (b). Here, the total refers
to the sum of the QCD, QED and their interference results.}
\label{fig:poladis}
\end{figure}

For the QCD contribution alone, we calculate the total cross sections
and $p_{t}$ distributions of $J/\psi$ production (dashed lines
in Fig.~\ref{fig:ptdis} (a)) and polarizations (the dashed lines in
Fig.~\ref{fig:poladis}) for the Tevatron and LHC. The results are in
agreement with the early
predictions\cite{Artoisenet:2007xi,Hagiwara:2007bq,Artoisenet:2008tc}
when the difference of the parameters are taken into account. And the
corresponding total cross sections are
\begin{subequations}
\begin{equation}
\sigma_{_{QCD}}(p\bar{p}\to J/\psi+c\bar{c}+X)=8.7
\,\mathrm{nb}\,(\mathrm{Tevatron}),
\end{equation}
\begin{equation}
\sigma_{_{QCD}}(p\bar{p}\to J/\psi+c\bar{c}+X)=169.6
\,\mathrm{nb}\,(\mathrm{LHC}).
\end{equation}
\end{subequations}

For the QED contribution alone, the total cross sections are 
\begin{subequations}
\begin{equation}
\sigma_{_{QED}}(p\bar{p}\to J/\psi+c\bar{c}+X)=52.3\,\mathrm{pb}\,(\mathrm{Tevatron}),
\end{equation}
\begin{equation}
\sigma_{_{QED}}(p\bar{p}\to J/\psi+c\bar{c}+X)=1.13\,\mathrm{nb}\,(\mathrm{LHC}).
\end{equation}
\end{subequations}
Since the integrated cross sections are mostly coming from the
contribution in small $p_{t}$ region, the total cross sections of
the QED part are about $(\alpha/\alpha_{s})^{2}\simeq1/100$ times
smaller than the QCD ones for both the Tevatron and LHC.

The $p_{t}$ distribution of $J/\psi$ QED production is shown in
Fig.~\ref{fig:ptdis} with lower (upper) dotted line for the Tevatorn
(LHC). It can be seen that the QED contribution is about 2 orders
of magnitude less than the QCD contribution at $p_{t}=3$ GeV and the
ratio of the QED contribution to the QCD one is $0.4~(0.7)$ at
$p_{t}=50~(100)$ GeV for the Tevatron (LHC). It means that the QED
contribution is indeed negligible in the small $p_t$ region, but is
important in the large $p_{t}$ region, especially at the LHC. The $p_t$
distributions of $J/\psi$ polarization from the QED part for the
Tevatron and LHC are plotted with the dotted-dashed lines in
Fig.~\ref{fig:poladis}. Both of the curves show that $J/\psi$ is
unpolarized in the small $p_{t}$ region and becomes increasingly
transverse in the region of large $p_{t}$. It is because the QED
part is dominated by the contribution from Type I diagrams in the small
$p_{t}$ region, while the Type II fragmentation contribution
dominates in the large $p_{t}$ region.

For the QED part alone, the contribution is
comparable to the QCD one in the large $p_{t}$ region and particularly
the polarization behavior of the former is very different from that
of the latter. Therefore, we also calculate the interference between
QED and QCD parts. It is found that the interference between the QED
and QCD parts of $pp(\bar{p})\to c\bar{c}[^3S_1,\underline{1}]+c\bar{c}$ is
positive and the relative phase angle is close to $\pi/2$.  The
total results are shown in Fig.~\ref{fig:ptdis} (b). The theoretical
uncertainties of the $p_t$ distribution for $J/\psi$ production are
studied by changing the renormalization and factorization
scale $\mu_f=\mu_r$ from $\mu_{0}/2$ to $2\mu_{0}$. They correspond
to the uncertainty bands of $J/\psi$ $p_{t}$ distribution
shown in Fig.~\ref{fig:ptdis} (b). 
The $p_{t}$ distribution of $J/\psi$ production is enhanced by $50\%$ compared to the
QCD result at $p_{t}=50$ GeV at the Tevatron. At the LHC, the
enhancement is 190\% at $p_{t}=100$ GeV. 
The $p_t$ distributions
of $J/\psi$ polarization are shown in Fig.~\ref{fig:poladis}. 
It is shown that the
polarization parameter $\alpha(p_{t})$ for the QCD part is almost
equals to zero in all regions at both the Tevatron and LHC. But it
becomes increasing slowly (see the solid line in
Fig.~\ref{fig:poladis}) when QED part is included in the full
calculation.

\section{Conclusion}
Until now, people still have not found a convincing mechanism to
explain the $J/\psi$ hadroproduction at the Tevatron. It is
suggested\cite{Artoisenet:2007xi} that the associated
production channel $J/\psi+c\bar{c}$, which may be observed through
measurement of the $J/\psi$ together with at least one charmed
hadron, may provide further insight to the mechanism responsible for
the $J/\psi$ hadroproduction. Both the color-singlet and color-octet
contributions for the associated $J/\psi$ production have been
calculated at $\alpha_{s}^{4}$ order in
Ref.\cite{Artoisenet:2008tc}. In this work, we have presented the
$\alpha^{2}\alpha_{s}^{2}$ order calculation for the associated
$J/\psi$ production at the Tevatron and LHC. Our calculation also
include the interference terms between the
$\alpha^{2}\alpha_{s}^{2}$ order QED part and $\alpha_{s}^{4}$ QCD
color-singlet part. Our results indicate the QED photon
fragmentation effect is very important, and it has a large impact on
$p_{t}$ distribution of $J/\psi$ production and polarization in
large $p_{t}$ region, especially for the LHC.

The QED part for $pp(\bar{p})\to J/\psi+c\bar{c}$
is also part of the NLO QCD corrections to photon fragmentation
inclusive $J/\psi$ production considered in our previous work~\cite{He:2009cq}. Both of them
scale like $1/p_{t}^{4}$, so unlike the $J/\psi$ QCD production case, the impact of the QED $J/\psi$
associated production contribution on the photon fragmentation inclusive $J/\psi$ production is not
very large for the $p_{t}$ distribution of $J/\psi$ production and polarization.

In this work, we do not study the $\Upsilon$ production associated
with the $b\bar{b}$ pair, though $\Upsilon$ can also be  produced by
photon fragmentation. There are two reasons. One is the electric
charge of bottom-quark is only half of that of charm-quark. It will
provide an additional $(1/2)^{4}$ suppression factor compared to the 
charm-quark case. The other is the mass of $\Upsilon $, which is about
3 times larger than the $J/\psi$ mass, which makes the photon
fragmentation effect for $\Upsilon$ not as outstanding as that
for $J/\psi$ case.

This work is supported by the National Natural Science Foundation of
China under Contract No.10775141 and by the Chinese Academy of Science under
Project No. KJCX3-SYW-N2.

\end{document}